\documentclass[sigconf]{acmart}

\copyrightyear{2024}
\acmYear{2024}
\setcopyright{acmcopyright}

\acmBooktitle{The IDE Workshop, April 14-20, 2024, Lisbon, Portugal}
\acmPrice{15.00}
\acmDOI{10.1145/3524459.3527349}
\acmISBN{978-1-4503-9285-3/22/05}

\usepackage{todonotes}

\usepackage{csquotes}
\usepackage{amsmath}
\usepackage{amsthm}
\usepackage{thmtools}
\usepackage{nicematrix}
\usepackage{multirow}
\usepackage{colortbl}
\usepackage{fontawesome5}
\usepackage{pifont}
\usepackage{cleveref}

\newtheoremstyle{recstyle}%
{0pt}
{0pt}
{}
{}
{}
{}
{ }
{\textbf{\thmname{#1}~\thmnumber{#2}: \thmnote{#3}}}
\theoremstyle{recstyle}
\newtheorem{recommend}{Recommendation}[subsection]
\Crefname{recommend}{recommendation}{recommendations}

\acmConference[The IDE Workshop, ICSE 2024]{The 45th International Conference on Software Engineering}{April 14–20, 2024}{Lisbon, Portugal}

\AtBeginDocument{%
  \providecommand\BibTeX{{%
    \normalfont B\kern-0.5em{\scshape i\kern-0.25em b}\kern-0.8em\TeX}}}

\copyrightyear{2024}
\acmYear{2024}
\setcopyright{acmlicensed}\acmConference[IDE '24]{2024 First IDE Workshop}{April 20, 2024}{Lisbon, Portugal}
\acmBooktitle{2024 First IDE Workshop (IDE '24), April 20, 2024, Lisbon, Portugal}
\acmDOI{10.1145/3643796.3648448}
\acmISBN{979-8-4007-0580-9/24/04}

\newcommand{\etal}{\textit{~et~al.}}

\begin{document}

\title{On the Integration of Spectrum-Based Fault Localization Tools into IDEs}

\author{Attila Szatmári}
\affiliation{%
  \institution{Department of Software Engineering, University of Szeged}
  \city{Szeged}
  \country{Hungary}
}
\email{szatma@inf.u-szeged.hu}

\author{Qusay Idrees Sarhan}
\affiliation{%
  \institution{Department of Computer Science, University of Duhok}
  \city{Duhok}
  \country{Iraq}
}
\email{qusay.sarhan@uod.ac}

\author{Gergő Balogh}
\affiliation{%
  \institution{Department of Software Engineering, University of Szeged}
  \city{Szeged}
  \country{Hungary}
}
\email{geryxyz@inf.u-szeged.hu}

\author{Péter Attila Soha}
\affiliation{%
  \institution{Department of Software Engineering, University of Szeged}
  \city{Szeged}
  \country{Hungary}
}
\email{psoha@inf.u-szeged.hu}

\author{Árpád Beszédes}
\affiliation{%
  \institution{Department of Software Engineering, University of Szeged}
  \city{Szeged}
  \country{Hungary}
}
\email{beszedes@inf.u-szeged.hu}

\newcommand{\RQ}[2]{\noindent\textbf{RQ#1:} \textit{#2}\smallskip\\}

\begin{abstract}
Spectrum-Based Fault Localization (SBFL) is a technique to be used during debugging, the premise of which is that, based on the test case outcomes and code coverage, faulty code elements can be automatically detected.
SBFL is popular among researchers because it is lightweight and easy to implement, and there is a lot of potential in it when it comes to research that aims to improve its effectiveness.
Despite this, the technique cannot be found in contemporary development and debugging tools, only a handful of research prototypes are available.
Reasons for this can be multiple, including the algortihms' sub-optimal effectiveness and other technical weaknesses.
But, also the lack of clear functional and non-functional requirements for such a tool, either standalone or integrated into IDEs.
In this paper, we attempt to provide such a list in form of recommendations, based on surveying the most popular SBFL tools and on our own researchers' and tool builders' experience.
\end{abstract}



\maketitle

\section{Introduction}

Spectrum-Based Fault Localization (SBFL) is a lightweight debugging technique that profiles the program's execution through its test cases, collects code coverage and test results (collectively called the spectrum), and then, based on statistics of failing and passing test cases and their relation to code elements, assigns suspiciousness scores to each such element, a statement or function in particular~\cite{survey1,survey2,intro2}.
We provide a brief overview of the technique in \Cref{SBFL_Background}.

Although it is a well-researched topic, SBFL is not widely adopted in the industry, and the research community is mostly aware of reasons why SBFL can be challenging. 
Sometimes meeting technical requirements can be difficult, and due to its statistical nature, SBFL can occasionally mislead developers during the debugging process leading to the rejection of the technique by practitioners~\cite{8009915,Kochhar2016Practitioners}.
Sarhan and Besz\'edes identified additional challenges in SBFL~\cite{SaB22a}.

In this paper, we concentrate on the aspects of implementing SBFL techniques into practical tools, in particular their integration into Integrated Development Environments (IDEs).
Hence, we do not deal with any other technical or theoretical aspects of the technique.
With this, we distinguish our study from a more conceptual comparison of related tools such as the one performed by Archana and Agarwal~\cite{10.1145/3511430.3511470}.

We base our investigation on three sources.
First, we perform a lightweight survey of existing SBFL tools to identify their main properties, and more importantly, the existing strengths and weaknesses (\Cref{Related_Works}).
Second, we summarize our findings from a recent empirical study that we performed with actual users who have been given a debugging task to be performed in an IDE with some of our own SBFL tools (\Cref{sec:process}).
Third, we collect our own experience as researchers working in the SBFL domain for several years and as tool builders who have experience in integrating debugging tools into various IDEs including Eclipse and JetBrains.

We list our findings in \Cref{Recomendations} from three perspectives: users', developers, and researchers', with the hope that future tool builders in this area could benefit from it when designing new tools, preferably integrated into IDEs.

\section{Spectrum-Based Fault Localization}
\label{SBFL_Background}
SBFL is a lightweight statistical fault localization approach in which every element (such as a statement or a function) will be assigned a suspiciousness score (usually between 0 and 1), indicating the possibility that the fault is caused by that element~\cite{Abreu2007}.
This score is then used to provide a ranked list of potentially faulty code elements to the user who can rely on this information during debugging.

To calculate these scores, SBFL operates on the so-called program spectrum. It consists of a coverage matrix and a test result vector.
In the most common case, the rows of the coverage matrix denote test cases of the program and the columns represent the elements of the source code. A cell value, $c_{i,j}$ is 1 if during the execution of the $i$\textsuperscript{th} test case the $j$\textsuperscript{th} element was executed, and 0 otherwise.
The test result vector ($e$) represents the results of the executions of the test cases, where $e_i$ is equal to 1 if the $i$\textsuperscript{th} test case failed and 0 otherwise (\Cref{fig:spectra}).

The next step is to calculate suspiciousness scores, which is usually done based on four basic metrics, where $\operatorname{metric~}_{\text{exec},\text{res}}(j)$ is the number of runs when the $j$\textsuperscript{th} element was executed ($\text{exec} = 1$) or not ($\text{exec} = 0$) and whether the inspected test case failed ($\text{res} = 1$) or passed ($\text{res} = 0$).
These four basic metrics are used to compose various suspiciousness formulas, such as Tarantula~\cite{tarantula_Sbfl}, Ochiai~\cite{ochiai_sbfl}, or Barinel~\cite{Barinel_Sbfl}.

Researchers advised many different types of extensions to this basic approach, including program slicing~\cite{Wen12:SBFLBasedOnProgramSlicing}, adding contextual information~\cite{DESOUZA18:ContextualizingSBFL}, user feedback~\cite{Gong12:InteractiveFL}, and machine learning~\cite{Gao18:MLforFL}, to enhance the effectiveness.

\begin{figure}
    \small
    \begin{NiceMatrixBlock}[auto-columns-width]
    \[
        \begin{array}{cccc}
            & n\text{ source code items} & & m\text{ tests} \\
            C = &
            \begin{bNiceMatrix}
                c_{1,1} & c_{1,2} & \cdots & c_{1,n} \\
                c_{2,1} & c_{2,2} & \cdots & c_{2,n} \\
                \vdots & \vdots & \ddots & \vdots \\
                c_{m,1} & c_{m,2} & \cdots & c_{m,n} \\
            \end{bNiceMatrix} &
            e = &
            \begin{bNiceMatrix}
                e_{1} \\
                e_{2} \\
                \vdots \\
                e_{m}
            \end{bNiceMatrix}
            \vspace{3mm}
            \\\\
            & 4\text{ basic metrics} \\
            \operatornamewithlimits{metric}\limits_{\text{exec},\text{res}}(j) = &
            \left|\left\{i\;|\;M_{i,j}=\text{exec} \land e_i = \text{res}\right\}\right| &
            \multicolumn{2}{c}{\text{exec},\text{res}\in \{0,1\}}
            \vspace{3mm}
            \\\\
            & n\text{ score values} \\
            s = &
            \begin{bNiceMatrix}
                s_{1} & s_{2} & \cdots & s_{n}
            \end{bNiceMatrix}    
        \end{array}
    \]
    \end{NiceMatrixBlock}

    \caption{Overview of SBFL}
    \label{fig:spectra}
\end{figure}

\section{SBFL Tools}
\label{Related_Works}
In this section, we provide an overview of the most frequently cited SBFL tools, highlighting their main features. It is worth noting that our research does not take into account non-functional requirements like speed or memory usage. Nevertheless, we acknowledge that these factors are crucial for boosting the acceptance of SBFL tools in the industrial sector, and we plan to address them in follow-up research.

Due to the nature of SBFL, it calls for its implementation into an IDE directly rather than as a standalone tool, but there are examples for both.

Wang \etal{}~\cite{FLAVS} proposed a fault localization tool named \enquote{FLAVS} for developers utilizing the Microsoft Visual Studio platform. The tool collects program spectrum information during program execution to identify suspicious program elements. It monitors all environmental parameters of the running program, such as thread counts, CPU use, and memory usage. Chen and Wang~\cite{UnitFL} have expanded the features of \enquote{FLAVS} to another tool named \enquote{UnitFL}. To shorten the program execution time, the tool makes use of program slicing.

A tool called \enquote{SFLaaS} was proposed by Sarhan\etal{}~\cite{SFLaaS} to locate faults in Python programs. It is offered as a cloud-based service instead of a plugin or command-line tool.
The program offers various tie-breaking techniques, approximately 80 SBFL formulas, the ability for the users to define their own formulas, and the display of code elements in various colors according to their suspiciousness scores, from most suspicious (red) to not suspicious (green).

For Java developers, Ribeiro\etal{}~\cite{Jaguar} proposed the SBFL tool \enquote{Jaguar}. The tool supports control and data flow, the two advanced spectra types. Moreover, it presents suspicious program elements graphically so that the user can quickly examine suspicious variables, statements, or methods. 

\enquote{Whyline} is a debugging tool for Java developers that was proposed by Ko and Myers~\cite{Whyline}. The tool utilizes slicing to aid developers in understanding the program behavior under test through a set of questions that are presented in a graphical and interactive way. The tool also logs program execution traces and the status of each program element, indicating its execution status. The tool allows users to view a program's execution trace and select an element at a specific execution point for examination.

The Eclipse plug-in tool \enquote{VIDA} was proposed by Hao\etal{}~\cite{VIDA} for Java programs. The tool provides the top ten most suspicious elements as possible breakpoints, along with a history of previous breakpoints and their current level of suspicion. Static dependency graphs are also produced to assist developers in comprehending the relationships between program elements.

A tool called \enquote{GZoltar} was proposed by Campos\etal{}~\cite{Gzoltar} to locate faults in Java programs, and it is provided as a plugin for the Eclipse IDE. It generates visually appealing and interactive diagnostic report visualizations like Sunburst and Treemap.

A tool called \enquote{CharmFL} was proposed by Sarhan\etal{}~\cite{CharmFL} to locate faults in Python programs using SBFL. The tool can be used as a standalone or as a plugin for the PyCharm IDE. It supports different code coverage types (i.e., statements, methods, and classes) with the possibility to investigate these types in a hierarchical approach.

The \enquote{CharmFL} tool has been extended in~\cite{Interactive_CharmFL} to support interactivity in fault localization. With the help of this feature, users can provide their feedback on the suspicious elements and help reorder them, which may improve the fault localization process. This is achieved by utilizing contextual information (i.e., function call graphs) during a failed test. As a result, users are able to examine the data flow traces of any failed test and pinpoint the fault's location more precisely. This interactivity feature is also provided by \enquote{iFL4Eclipse} an Eclipse plug-in SBFL tool~\cite{using-contextual-knowledge-in-interactive-fault-localization-ferenc-horvath-arpad-beszedes-bela-vancsics-gergo-balogh-laszlo-vidacs-tibor-gyimothy}.

\section{Experiences from a User Study}
\label{sec:process}

In a recent study, empirical experiments were conducted to assess the effectiveness of the SBFL technique with user feedback integrated into an IDE~\cite{using-contextual-knowledge-in-interactive-fault-localization-ferenc-horvath-arpad-beszedes-bela-vancsics-gergo-balogh-laszlo-vidacs-tibor-gyimothy}.
The research was performed on different levels; with simulation, using students, and with professional developers.
Among other things, conclusions could be drawn from this study regarding the user's perception of the tool used, so this information is very useful for our purposes in the present paper.

We reuse this data from the user study, in particular, the participants' oral responses in form of transcripts of the experiment sessions.
For constructing the transcripts, a \emph{think-aloud technique} was utilized which was performed with six professional developers. It involved 24 debugging sessions, with each of the developers participating in four of them.
Their task was to find and fix a bug in existing software using an SBFL tool plug-in in the Eclipse IDE.
During the process, their ways of thinking were recorded, which included navigating and understanding the code, and in particular using the tool's functions.

In the original paper by Horváth \etal{}~\cite{using-contextual-knowledge-in-interactive-fault-localization-ferenc-horvath-arpad-beszedes-bela-vancsics-gergo-balogh-laszlo-vidacs-tibor-gyimothy}
the main goal was to evaluate the effectiveness of interactive FL in a variety of scenarios. 
The tasks for the experiments were selected from various projects by choosing bugs that did not require specific domain knowledge.
This approach was intended to simulate situations where developers debug an unfamiliar program, which is common in real-world software development.
The participant pool included undergraduate and graduate students as well as professional developers, aiming to cover a diverse range of experience levels.
The experimental setup was balanced in terms of participant distribution across different groups.
The complexity of tasks and potential prior knowledge of the participants about projects or bugs was considered during the design phase.

We processed the transcript of these sessions with a focus on the participants' opinions about the tools.
To enhance objectivity, we instructed the ChatGPT 4.0 LLM engine~\cite{large-language-models-understand-and-can-be-enhanced-by-emotional-stimuli-cheng-li-jindong-wang-yixuan-zhang-kaijie-zhu-wenxin-hou-jianxun-lian-fang-luo-qiang-yang-xing-xie} to extract all the expectations (both implicit and explicit) of the software engineers from the transcripts. Implicit expectations refer to improvements in the IDE and plug-ins that make their tasks easier, which can be inferred from their actions and behavior. On the other hand, explicit expectations are specific requests for features or explicitly expressed wishes by the developers.


\begin{figure}[h]
\centering
\includegraphics[width=0.9\columnwidth]{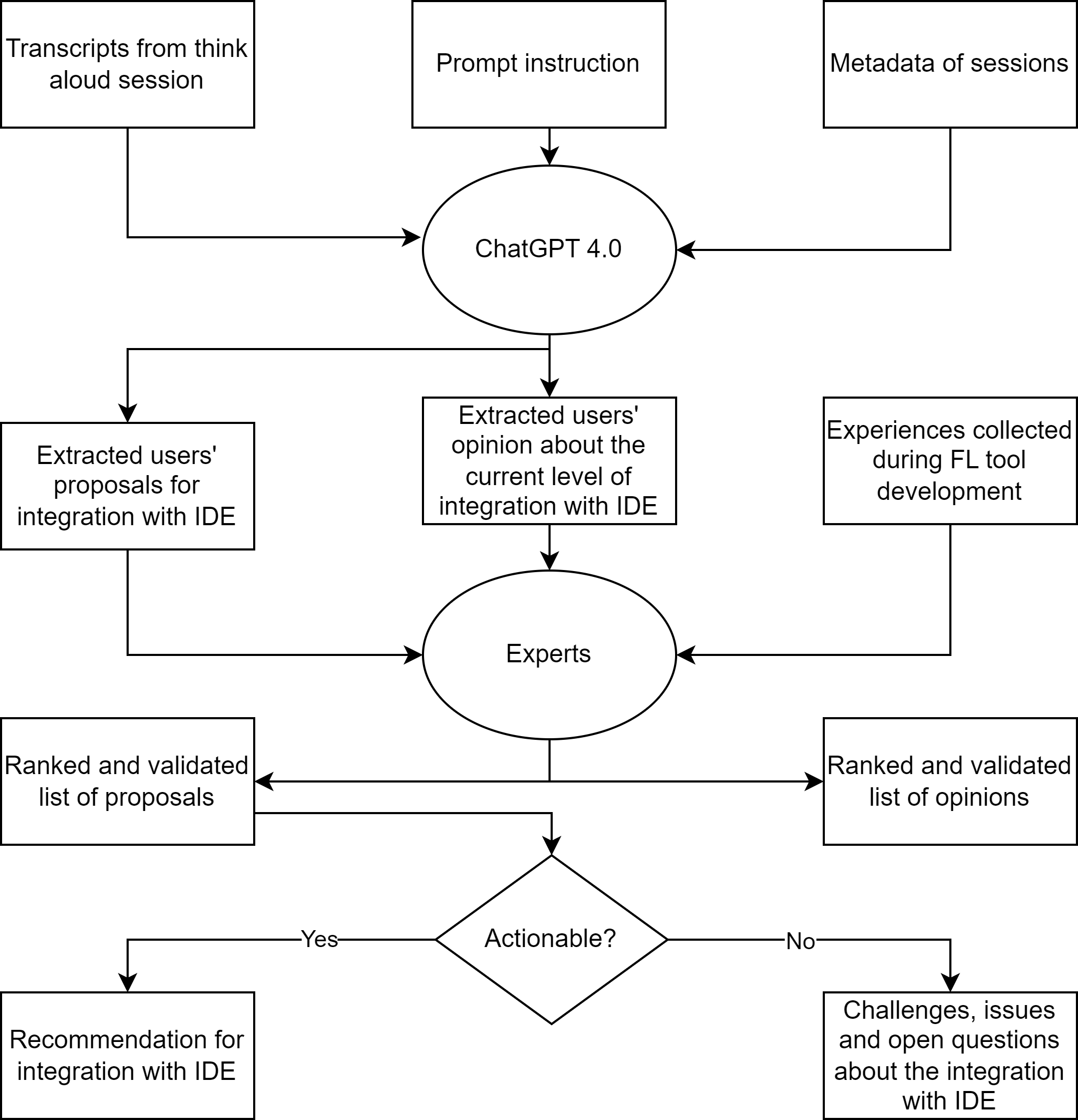}
\caption{Recommendation extraction process}
\label{fig:methodology}
\end{figure}

The recommendation extraction process is shown in \Cref{fig:methodology}.
The transcripts of the think-aloud sessions, prompt instructions and metadata from the previous experiment are entered into ChatGPT to help process and summarize these diverse data sources. 
All prompts are available in the online appendix of this paper~\cite{Szatmari23:OnlineAppendix}.
Users' proposals for integration to the IDE and their opinions on the current level of integration summarized by ChatGPT were then carefully investigated by the authors of this paper.
We treat this step as an expert evaluation as we could contribute valuable insights based on our experience with the development of SBFL tools. 
We then constructed a list of proposals and opinions. 
Depending on their actionability, recommendations for integration with the IDE are formulated or questions are raised that address open challenges and issues. 

ChatGPT extracted more than 150 expectations from over 250 thousand tokens' transcriptions. We utilized a tool to filter semantic duplicates, which resulted in 17 distinct raw expectations. 
To validate the raw expectations, we used the Elo rating system~\cite{elo-rating-as-a-tool-in-the-sequential-estimation-of-dominance-strengths,the-proposed-uscf-rating-system--its-development-theory--and-applications}, typically used for calculating the relative skill levels of players in zero-sum games like chess. In this case, each expectation was considered a \enquote{player} competing for the votes of software engineering experts who had to decide which expectation was more important during debugging or fault localization. To determine the winner, we counted the votes for each expectation and updated their Elo rating accordingly.
We used two free scaling parameters, K and c, with their typical values of K=32 and c=400. Each expectation had an initial rating of 1500. Our matching strategy combined elements of randomness, exploration, and exploitation to ensure a balanced and fair evaluation of all items. The source code of our rating strategies is available in the online appendix~\cite{Szatmari23:OnlineAppendix}.


In the final phase, we compared the expectations that were ranked with the features provided by the fault localization tools. This helped us create the final recommendations and assemble the list of open questions, to be discussed in \Cref{Recomendations}.

\section{Recommendations}
\label{Recomendations}


In this section, we summarize and blend our findings from the tool survey in \Cref{Related_Works}, the user study from \Cref{sec:process}, and our own experiences.
First, we give an overall list of expectations we identified and assess the investigated tools from the perspective of how they implement them in \Cref{tab:expectations}.
In the last part of this section, we summarize what we believe are the most important recommendations from the users', developers', and researchers' perspectives, respectively.

\newcommand{\fade}[1]{\textcolor{gray}{#1}}
\newcommand{\head}[1]{\textbf{#1}}
\newcommand{\noOneDone}{\rowcolor{red!8}\fade{none}}
\newcommand{\fewDone}{\rowcolor{blue!15}\fade{few}}
\newcommand{\GENERALresponsible}{\raisebox{.3ex}{{\tiny\faIcon{globe}}}}
\newcommand{\IDEresponsible}{\raisebox{.3ex}{{\tiny\faIcon{sitemap}}}}
\newcommand{\PLUGINresponsible}{\raisebox{.25ex}{{\tiny\faIcon{plug}}}}
\newcommand{\none}{\fade{\O}}
\newcommand{\noneLine}{\none &\none &\none &\none &\none &\none &\none &\none &\none &\none &\none &\none}

\begin{table*}[t]
    \centering
    \label{tab:expectations}    
    \caption{
        Summary of expectations and their relationship to existing SBFL tools.
        \\{
            \normalfont\footnotesize C = CharmFL, F = FLAVS, G = GZoltar, J = Jaguar, S = SFLaaS, U = UnitFL, V = VIDA, W = Whyline, i = iFL4Eclipse;
            \colorbox{red!8}{red} = no tool implements or plans it, \colorbox{blue!15}{blue} = less then 3 tools (1\textsuperscript{st} quartile) implement or plan it;
            \GENERALresponsible{} = responsibility is unassigned; \IDEresponsible{} = IDE is responsible; \PLUGINresponsible{} = plug-in is responsible;
            Responsibilities were determined based on our experiences, taking into account expectations, their manifestations in tools, and tools' visions.
        }
        }
    \resizebox{\textwidth}{!}{
\begin{tabular}{@{}clr|ll|llll|ll|llll@{}}
& & & \multicolumn{2}{c}{\head{Implementation}} & \multicolumn{4}{c}{\head{Validation}} & \multicolumn{2}{c}{\head{Maturity}} & \multicolumn{4}{c}{\head{Ecosystem}}\\
& \head{Short Summary} & \multirow{2}{*}{\head{Rating}} & \multirow{2}{*}{\head{Implemented}} & \multirow{2}{*}{\head{Planned}} & \multirow{2}{*}{\head{None}} & \head{by} & \head{with} & \head{with} & \head{Alpha} & \head{Beta} & \head{Stand-} & \multirow{2}{*}{\head{Eclipse}} & \head{Visual} & \multirow{2}{*}{\head{PyCharm}}\\
& & & & & & \head{Authors} & \head{Students} & \head{Prof.} & \head{Version} & \head{Version} & \head{alone} & & \head{Studio}\\
\hline
& code navigation and search & 1.00 & 6 \fade{(C\IDEresponsible;F\IDEresponsible;G\IDEresponsible;J\IDEresponsible;U\IDEresponsible;i\IDEresponsible)} &  & 3 \fade{(F;G;U)} &  & 2 \fade{(J;i)} & 1 \fade{(i)} & 1 \fade{(C)} & 5 \fade{(F;G;J;U;i)} &  & 3 \fade{(G;J;i)} & 2 \fade{(F;U)} & 1 \fade{(C)}\\
\fewDone & clearer error messages & 0.90 & 1 \fade{(G\GENERALresponsible)} &  & 1 \fade{(G)} &  &  &  &  & 1 \fade{(G)} & & 1 \fade{(G)} &  & \\
\fewDone & tests' running optimization & 0.88 &  & 2 \fade{(G\GENERALresponsible;J\GENERALresponsible)} & 2 \fade{(G;J)} &  &  &  &  & 2 \fade{(G;J)} & & 2 \fade{(G;J)} &  & \\
\fewDone & code execution path visualization & 0.88 & 1 \fade{(J\GENERALresponsible)} &  &  1 \fade{(J)} &  &  &  &  & 1 \fade{(J)} & & 1 \fade{(J)} &  & \\
\fewDone & clearer debugging process & 0.86 & 2 \fade{(V\GENERALresponsible;W\GENERALresponsible)} &  & 1 \fade{(V)} &  & 1 \fade{(W)} & 1 \fade{(W)} & 1 \fade{(V)} & 1 \fade{(W)} &  1 \fade{(W)} & 1 \fade{(V)} &  & \\
& intuitive UI & 0.86 & 7 \fade{(C\GENERALresponsible;F\GENERALresponsible;G\GENERALresponsible;J\GENERALresponsible;S\GENERALresponsible;U\GENERALresponsible;W\GENERALresponsible)} &  & 5 \fade{(F;G;J;S;U)} & 1 \fade{(C)} & 2 \fade{(W;i)} & 2 \fade{(W;i)} & 2 \fade{(C;S)} & 5 \fade{(F;G;J;U;W)} & 1 \fade{(W)} & 1 \fade{(G;J)} & 1 \fade{(F;U)} & 1 \fade{(C)}\\
\fewDone & advanced debugging tools & 0.83 & 1 \fade{(W\GENERALresponsible)} &  &  &  & 1 \fade{(W)} & 1 \fade{(W)} &  & 1 \fade{(W)} & 1 \fade{(W)} &  &  & \\
& responsive IDE & 0.81 & 3 \fade{(C\GENERALresponsible;J\IDEresponsible;F\IDEresponsible;G\GENERALresponsible)} &  & 3 \fade{(C;G;J)} &  &  &  & 1 \fade{(C)} & 2 \fade{(F;G)} & & 1 \fade{(G)} & 1 \fade{(F)} & 1 \fade{(C)}\\
\noOneDone & customizable interface & 0.79 & \noneLine{} \\
& intuitive code coverage visualization & 0.78 & 2 \fade{(C\GENERALresponsible;G\GENERALresponsible)} & 1 \fade{(J\GENERALresponsible)} & 3 \fade{(C;G;J)} &  &  &  & 1 \fade{(C)} & 2 \fade{(G;J)} & & 2 \fade{(G;J)} &  & 1 \fade{(C)}\\
& streamlined test processes & 0.77 & 8 \fade{(C\PLUGINresponsible;F\PLUGINresponsible;G\PLUGINresponsible;J\PLUGINresponsible;S\PLUGINresponsible;U\PLUGINresponsible;V\PLUGINresponsible;i\PLUGINresponsible)} &  & 7 \fade{(C;F;G;J;S;U;V)} & &  & 1 \fade{(i)} & 3 \fade{(C;S;V)} & 5\fade{(F;J;G;U;i)} & 1 \fade{(S)} & 4 \fade{(G;J;V;i)} & 2 \fade{(F;U)} & 1 \fade{(C)}\\
& IDEs' compatibility with plugins & 0.70 &  \fade{(C\IDEresponsible;G\IDEresponsible;U\IDEresponsible)} &  &  \fade{(C;G;U)} &  &  &  & 1 \fade{(C)} & 2 \fade{(G;U)} &  & 2 \fade{(G;U)} &  & 1 \fade{(C)}\\
\noOneDone & code changes' error message clarity & 0.69 & \noneLine{} \\
& plugin integration and interaction & 0.61 & 3 \fade{(C\GENERALresponsible;G\GENERALresponsible;U\GENERALresponsible)} &  & 3 \fade{(C;G;U)} &  &  &  & 1 \fade{(C)} & 2 \fade{(G;U)} &  & 1 \fade{(G)} & 1 \fade{(F)} & 1 \fade{(C)}\\
\noOneDone & integrated domain knowledge & 0.60 & \noneLine{} \\
\noOneDone & integrated documentation & 0.00 & \noneLine{} \\
\hline
& interactivity between SBFL and developers & - & 3 \fade{(C\PLUGINresponsible;V\PLUGINresponsible;i\PLUGINresponsible)}&1\fade{(S\PLUGINresponsible)} & 3 \fade{(C;S;V)}& & & 1 \fade{(i)}& 3 \fade{(C;S;V)} & 1 \fade{(i)} & 1 \fade{(S)}& 2 \fade{(V;i)} & & 1 \fade{(C)}\\ 
\fewDone & hierarchical view based on code structure & - & 3 \fade{(C\GENERALresponsible;G\GENERALresponsible;J\GENERALresponsible)} & & 2 \fade{(C;G)}& &1 \fade{(J)} & & 1 \fade{(C)}&2 \fade{(G;J)}& &2 \fade{(G;J)}& & 1 \fade{(C)}\\ 
\fewDone & accessibility-focused UIs & - &1 \fade{(G\GENERALresponsible)} & & 1 \fade{(G)}& & & & 1 \fade{(G)}& & & 1 \fade{(G)} & & \\ 
\noOneDone & enabling teamwork and sharing & - & \noneLine{} \\ 
\fewDone & inspect suspicious items from different aspects & - & 1 \fade{(J\GENERALresponsible)} & & & &1 \fade{(J)} & & & 1 \fade{(J)}& & 1 \fade{(J)} & & \\ 
\fewDone & optimize SBFL with test selection \& prioritization & - & 1 \fade{(V\GENERALresponsible)} & & 1 \fade{(V)}& & & & 1 \fade{(V)}& & & 1 \fade{(V)} & & \\ 
\noOneDone & incorporate AI into SBFL & - & \noneLine{} \\ 
\noOneDone & handling code element context switching & - & \noneLine{} \\ 
\noOneDone & on-demand reanalysis upon code change & - & \noneLine{} \\ 
& suspiciousness score visualization & - & 5 \fade{(F\GENERALresponsible;G\GENERALresponsible;J\GENERALresponsible;U\GENERALresponsible;V\GENERALresponsible)} &1\fade{(S\PLUGINresponsible)} &5 \fade{(F;G;S;U;V)} & & 1 \fade{(J)}& & 2 \fade{(S;V)}& 4 \fade{(F;G;J;U)} & 1 \fade{(S)} & 3 \fade{(G;J;V)} & 2 \fade{(F;U)} & \\ 
\noOneDone & understanding the tool's reasoning & - & \noneLine{} \\ 
\noOneDone & (smart) customizable feature set & - & \noneLine{} \\ 
\fewDone & customizable SBFL algorithm & - &2 \fade{(F\PLUGINresponsible;S\PLUGINresponsible)} & &2 \fade{(F;S)} & & & & &2 \fade{(F;S)} & 1 \fade{(S)}& & 1 \fade{(F)} & \\ 
\end{tabular}
    }
\end{table*}

\Cref{tab:expectations} is composed of two parts: expectations from the user study in the upper half and other generic expectations based on the tool survey and the authors' own experiences in the lower half.
The first column is a short summary of the expectation with the associated Elo rating in the second column (for user study only).
Columns 3 and 4 indicate if the expectation is implemented or planned in the respective tools, columns 5--8 show how the feature has been validated, columns 9 and 10 estimate the tool's maturity, and the last 4 columns show how the tool is implemented, either a standalone or an IDE plugin.


Note that since we identified a relatively large number of expectations (column one of \Cref{tab:expectations}), due to space limits we cannot explain all in detail;
their description can be found in the online appendix~\cite{Szatmari23:OnlineAppendix}.

\subsection{Users perspective}
Based on the data presented in \Cref{tab:expectations}, we identified two interwoven trends in user expectations. First, users expect SBFL tools to be seamlessly integrated in their workflow and efficient. Second, users prefer these tools to be easy to understand and use. 

\begin{recommend}[Seamless integration into IDEs and developers workflow.]
For developers, IDEs are considered complete tools that
help them in their daily workflow rather than just collections of plug-ins. They expect the UI to be intuitive, along with various visualizations such as coverage and suspiciousness score. 
The plug-ins should work cohesively with each other and with the IDE to meet the users' expectations.
\end{recommend}

\begin{recommend}[Transparent reasoning behind the tool.]
As problem solvers, developers love to automate complex processes. However, they may not fully trust the SBFL algorithms due to the lack of familiarity. To earn their trust, SBFL tools should provide clear error messages, easily accessible documentation, explainable operation, and drill-down views that help developers understand the reasoning behind the tool's suggestions.
\end{recommend}


\subsection{Developers perspective}
In this section, we provide advice on how to incorporate SBFL tools into IDEs, based on our experience as developers.

\begin{recommend}[Find efficient ways to produce, store and process runtime data.]
    The SBFL spectrum, including test case execution information, should be kept in a data structure that allows for quick access to its individual elements.
    This is essential if we want to ensure efficient operation with a responsive user interface of the tool.
\end{recommend}

\begin{recommend}[Programmers should strive to fully utilize IDEs' data and API features.]
The IDE should take charge of certain components of the SBFL process, and the developer should work together with the IDE to make sure that the process is handled properly and minimize solutions outside of the IDE.
For instance, it is recommended that test execution is done directly within the IDE.
\end{recommend}

\begin{recommend}[Ensure that code elements are easily identifiable.]
IDEs store program execution data in their own format, which may differ from the static source code structure. SBFL tools require both dynamic analysis and static source code representation, especially when working with loosely typed scripting languages like Python or JavaScript.
Information from static and dynamic analysis is often a challenge to match.
\end{recommend}

\begin{recommend}[Reporting about validation efforts.]
During the inspection of the selected set of tools against the expectations collected from users and developers, we noticed a concerning trend regarding the authors' validation efforts. 
Although most of the papers did a good job of reporting the features and methods underlying their research, in many cases we were unable to find concrete evidence of how the authors validated their contributions, and we believe this should be improved. 
\end{recommend}

\begin{recommend}[Decide on the responsibility of tool and IDE developers.]
The developers of tools and IDEs should put into practice the expectations outlined in \Cref{tab:expectations}. However, it is not always obvious who should be responsible for what.
Also, it is not always reasonable to expect IDE developers to respond to a feature request made by a plug-in developer.
For instance, implementing IDE specific features such as enabling code navigation and search within the codebase should be the responsibility of the IDE developers.
But implementing a new, hierarchical code visualization, for instance, is probably the tool-builder's job.
To allow easier assignment of responsibility, we may consider the following questions. \emph{\enquote{Can the plugin implement the expectation without the help of the IDE?}} \emph{\enquote{Does the plugin fulfill its goal without the given expectation?}} \emph{\enquote{Are the IDE and plugin necessary to tightly cooperate to implement the expectation?}}
\end{recommend}

\subsection{Researchers perspective}
Although SBFL has extensive literature, active work continues with new formula exploration, adding context to the spectrum, and developing different score calculation algorithms including AI-enabled approaches, to name a few. Researchers can take steps to prepare for different types of follow-up research.

\begin{recommend}[Make data computed by SBFL tools easily available for researchers.]
One way to do this is to provide dedicated logging and dumping mechanisms to retrieve data from the tools. This can be useful for analyzing users' behavior as well.
\end{recommend}

\begin{recommend}[Make SBFL tools' underlying logic customizable.]
Another way is to offer customization options within the tool itself. This can be achieved through settings, a modular architecture, or simply by making the tool open-source. While expecting researchers to include these customization systems out-of-the-box may be unrealistic, researchers should not prevent others from building upon their work either.
\end{recommend}


\begin{recommend}[Easy implementation and experimentation with new algorithms.]
Devising new human-computer interactions for emerging scientific methods, like different flavors of SBFL, is presumably more suited for researchers than professional programmers. 
Hence, it is important for researchers to make it as easy as possible to implement new features and perform experimentation with prototype ideas.
\end{recommend}

\begin{recommend}[Contribute to open science.]
Finally, we highly encourage tool developers to contribute to open science wherever they can. 
Besides the recommendations we already mentioned above, this would mean making  the implementation, test data and other artifacts openly available.
A possible framework that could be followed in particular is the \textit{FAIR guideline for software}~\cite{Lamprecht20:TowardsFAIR}, which includes aspects like Findability, Accessibility, Interoperability and Reusability, among others.


\end{recommend}

\section{Threats to Validity}
Our study focused on integrating SBFL tools into IDEs. However, we understand that there are several critical factors that could impact the robustness and applicability of our findings. We acknowledge that the selection of specific tools and environments for this study was necessary, but it limits the breadth of our conclusions. Therefore, the insights gathered from this study may not be applicable to all varieties of these tools or different development environments. Future research could benefit from including a wider range of tools to broaden the scope of our findings.

It is important to recognize that the interpretation of feedback from users and developers is subjective and based on individual experiences and perceptions. Therefore, to minimize the impact of this subjectivity, future experiments should consider increasing the number of participants and implementing measures to ensure anonymity. 

The methodology used to assess the efficacy and usability of tools integrated within IDEs may not encompass every facet of the user and developer experience, which could result in a partial evaluation. Therefore, a more comprehensive investigation that solicits feedback from a diverse group of users is necessary to offer a more holistic understanding of the subject matter.

Finally, the rapidly evolving nature of SBFL tools and IDEs poses a formidable challenge to the long-term relevance of research findings. It is crucial to periodically re-evaluate the efficacy of these tools, taking into account the latest advancements, to ensure that our research remains relevant and cutting-edge. Such periodic assessments are crucial for maintaining the credibility and applicability of our research outcomes.

\section{Conclusions}
\label{Conclusions}

This paper is a work towards practically usable SBFL in IDEs.
Out of the total of 29 expectations, only 18 were implemented (62\%), which implies that 11 expectations were left unfulfilled. It is noteworthy that only one of the unimplemented expectations was planned. Hence, it is safe to assume that the acceptance of SBFL tools by industrial Integrated Development Environment (IDE) users is still far from being achieved.
The technique itself is promising since it could significantly help in semi-automated debugging, but currently the implementations are not mature enough to be considered by professional programmers.

Based on surveying existing tools, asking users' opinions, and our own experiences, we compiled a list of concrete recommendations for future tool builders for this task.
It is without doubt that the underlying technology should also be further researched to reach the efficiency and precision that are required in a practical, everyday context as well, not only in a lab setting.
However, our concerns regarding the implementation into IDEs should be equally considered, and parallel development of the two can hopefully give birth to usable SBFL plug-ins in some of the popular IDEs in the future.

\section*{Acknowledgements}

This research was carried out in project TKP2021-NVA-09 supported
by the Ministry of Innovation and Technology of Hungary from the
National Research, Development and Innovation Fund, financed
under the TKP2021-NVA funding scheme.

\bibliographystyle{ACM-Reference-Format}
\bibliography{main}

\appendix

\end{document}